\documentstyle[aps,epsf,epsfig,twocolumn,psfig,array]{revtex}

\begin{document}

\draft \preprint{}

\title{De Broglie Wavelength Reduction for a Multi-photon Wave Packet}
\author{O. Steuernagel}
\address{Dept. of Physical Sciences,
University of Hertfordshire, College Lane, Hatfield, AL10 9AB, UK}
\date{\today}
\maketitle
\begin{abstract}
An experiment is proposed that permits the observation of the reduced de Broglie wavelengths of
two and four-photon wave packets using present technology. It is suggested to use a Mach-Zehnder
setup and feed both input ports with light generated by a single non-degenerate down-conversion
source. The strong quantum correlations of the light in conjunction with boson-enhancement at
the input beam splitter allow to detect a two- and fourfold decrease in the observed de Broglie
wavelength with perfect visibility. This allows a reduction of the observed de Broglie
wavelength below the wavelength of the source.
\end{abstract}

\narrowtext



Recently Jacobson {\it et al.}~\cite{Jacobson} gave a theoretical description of how to measure
the effective de Broglie wavelength of a quantum system in a coherent many-particle state using
a scheme employing non-linear beam splitters. Their work concluded that many-particle systems
with a sufficiently sharply defined number of consti\-tu\-ents, in the best case prepared in a
pure number state~$| N \rangle $, should show a reduction from the single particle wavelength
$\lambda$ to the effective multi-particle wavelength $\lambda/N$. This is the case even if the
constituents are not bound together. For bound systems such as poly-atomic
molecules~\cite{pritchard,Zeilinger_nature} this fact is well known, but for unbound systems
such as $N$-photon Fock states~$| N \rangle $ this finding is perhaps surprising.

The scheme suggested by Jacobson {\it et al.} uses non-linear beam-splitters in order to emulate
an effective binding of the particles. In a two-mode interferometer this input beam-splitter
channels the unbound consti\-tu\-ents of the system such that they simultaneously follow the
'upper' interferometric channel~$u$ in superposition with all of them passing through the
'lower' channel~$l$; the quantum state inside the interferometer reads
%
\begin{eqnarray}
| \psi \rangle_{inside} = \frac{| N \rangle_u | 0 \rangle_l + | 0 \rangle_u | N \rangle_l
}{\sqrt{2}} \; . \label{best.state.inside}
\end{eqnarray}
%
Compared to the case of a bound system this state obviously corresponds to an $N$-atomic
molecule traversing the interferometer undivided, thus maximally shrinking the system's de
Broglie wavelength.

\begin{figure}
\epsfverbosetrue \epsfxsize=3.6in \epsfysize=1.3in
%
%
%
\epsffile[12 612 583 803]{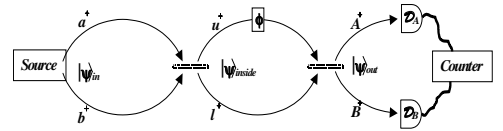}
\caption{Sketch of the interferometer: dotted lines outline balanced beam splitters, $\cal D$
stands for detectors and $a,b,u,l,A$ and $B$ label the modes before, inside and beyond the
interferometer. $\phi$ is the phase shifter in the upper channel. \label{figure1}}
\end{figure}

The output beam mixer of the interferometer, see Fig.~\ref{figure1}, is supposed to be of the
same non-linear kind as the input beam-splitter. Unfortunately, the use of non-linear
beam-splitters amounts to a considerable practical problem with regards to the experimental
implementation of the ideas presented in reference~\cite{Jacobson}: at present, neither the
required large magnitudes of the non-linearities nor their different orders for different
particle number $N$ are practically achievable.

Inspired by Jacobson's work, but following a rather different approach, Fonseca {\it et al.}
managed to measure the predicted halving of the wavelength for a two-photon state generated in
parametric down-conversion~\cite{Fonseca99}. The down-conversion source allowed them to abandon
the non-linear beam-splitter in the input of their Young's double-slit setup, and a two-particle
coincidence detection scheme allowed them to get rid of the second non-linear beam mixer. Here,
I want to present a modification to their approach (borrowing ideas from~\cite{Burnett,Ou99}
and~\cite{Ryff01}) that should allow for the observation of a fourfold reduction of the de
Broglie wavelength with currently available technology~\cite{Ou99,ghzexperiment}.

The key idea is to use linear beam-splitters in a Mach-Zehnder setup and utilize
boson-enhancement to generate good approximations to the ideal state~(\ref{best.state.inside})
inside the interferometer. Combined with suitable post-selection of detection events this
approach, in principle, even allows for the observation of more than fourfold reductions in the
measured de Broglie wavelength of unbound multi-particle systems.

So far, only variations with the wavelength of the pump-beam, i.e. halving of the effective de
Broglie wavelength (using one photon pair) have been
observed~\cite{Fonseca99,Horne89,RarityTapster,Herzog94}, consequently the idea presented here
will allow for a test of a quantum effect not observed before, namely, a reduction of the
multi-particle de Broglie wavelength below that of the generating
source~\cite{Ryff01,quartering}. See reference~\cite{Sackett00}, however, for the observation of
a quartering of the joint phase of the internal degrees of freedom of four entangled ions in a
trap.
%
%
%
%

From now on we will only discuss photons passing a Mach-Zehnder interferometer, see
Fig.~\ref{figure1}. Implicitly this discussion covers many bosonic systems and all types of
two-mode interferometers.

Rather than using non-linear beam splitters, even for two-photon states, it is easier to prepare
the desired interferometric state~(\ref{best.state.inside}) from photon-pairs generated in
type~$II$ (non-degenerate) parametric down-conversion. Fonseca {\it et al.} used a pump beam
with a momentum characteristic such that both photons would always pass the same
slit~\cite{Fonseca99}. Hence, the state $(| 1,1 \rangle_u | 0,0 \rangle_l + | 0,0 \rangle_u |
1,1 \rangle_l )/\sqrt{2}$ was created, where the two slots $| \cdot,\cdot \rangle$ refer to the
two orthogonal polarization modes of the photons and the indices $u$ and $l$ refer to either
slit in their double slit setup. This state is very similar to the desired
form~(\ref{best.state.inside}) and works just as well, as long as interferometric phase changes
apply to both polarization modes simultaneously. Experiment~\cite{Fonseca99} confirmed some
ideas presented in~\cite{Jacobson} and earlier results about joint measurements on
down-converted photons~\cite{Horne89,RarityTapster}.

Another recent proposal\cite{Ryff01} suggests to use a Mach-Zehnder layout and a projective
measurement involving a third auxiliary photon (from a second down-conversion photon pair) in
order to prepare state~(\ref{best.state.inside}) for one photon pair inside the interferometer.

But both these approaches do not scale favourably for higher photon numbers, in the latter
case~\cite{Ryff01} the use of auxiliary photons leads to extra channels and losses in the
interferometer. In the former case of experiment~\cite{Fonseca99} the momentum selection trick
cannot be extended to higher photon numbers because multiple pump photons would be involved;
this leads to a multiple product of the single-pair state mentioned above, i.e., in general to a
state very different from the desired form~(\ref{best.state.inside}). Therefore, another
ingredient is needed to channel the photons through the interferometer, for this,
boson-enhancement can be used.

As is known from the Mandel-dip experiment~\cite{Mandeldip,SalehTeich,Ryff01} the preparation of
state~(\ref{best.state.inside}) using boson-enhancement is straight\-for\-ward for two photons:
spontaneous parametric down-conversion generates single and multiple photon pairs in a two-mode
squeezed vacuum~\cite{Scully.buch} from the vacuum state $| 0 \rangle$
%
\begin{eqnarray}
| \psi \rangle_{in} = \frac{1}{\sqrt{1-|\alpha|^2}}
 \sum_{n=0}^{\infty} \frac{(\alpha \; a^\dagger b^\dagger)^n}{n!} \, | 0 \rangle \; ,
\label{down.conversion.state}
\end{eqnarray}
%
here $a^\dagger$ and $b^\dagger$ are the bosonic creation operators for the field modes of the
interferometer input ports, see Fig.~\ref{figure1}. With probability $|\alpha|^2/(1-|\alpha|^2)$
a suitable array of lossless detectors~\cite{lossless}, operating in coincidence, will detect
two photons. This allows us to retrodict that state~(\ref{down.conversion.state}) became
projected into state $|1,1\rangle$ which entered the interferometer and became converted into
the state
%
\begin{eqnarray}
| \psi \rangle_{inside} = \frac{| 2 \rangle_u | 0 \rangle_l + | 0 \rangle_u | 2 \rangle_l
}{\sqrt{2}} \; , \label{state.2.inside}
\end{eqnarray}
%
where the transformations for a balanced beam-splitter
%
\begin{eqnarray}
a^\dagger = ({u^\dagger + i \, l^\dagger})/{\sqrt 2}, \quad b^\dagger = ({u^\dagger - i \,
l^\dagger})/{\sqrt 2}
 \label{mode.transformation}
\end{eqnarray}
%
have been assumed~\cite{SalehTeich}. The behaviour of such photon pairs has been studied
intensely~\cite{Fonseca99,RarityTapster,Herzog94,Mandeldip}.

For four and more photons, our scheme only allows us to prepare states similar to the ideal
state~(\ref{best.state.inside}). Let us assume that a fourfold coincidence has been detected
(see~\cite{Ou99,ghzexperiment}), this occurs with probability $|\alpha|^4/(1-|\alpha|^2)$. It
allows us to infer that the incident state was of the form $| \psi \rangle_{in}=a^{\dagger 2}
b^{\dagger 2} | 0 \rangle /2$. Inside the interferometer this becomes~\cite{Ou99} $(\varepsilon
\equiv 1)$
%
\begin{eqnarray}
| \psi \rangle_{inside} & = & \left( \frac{u^{\dagger\, 4}}{8} + \frac{u^{\dagger\, 2}
l^{\dagger\, 2}}{4} + \frac{l^{\dagger\, 4}}{8} \right) | 0 \rangle
\\
& = & \sqrt{\frac{3}{4}} \left( \frac{| 4 \rangle_u | 0 \rangle_l + | 0 \rangle_u | 4
\rangle_l}{\sqrt 2} \right) + \frac{\varepsilon}{\sqrt{4}} \, | 2 \rangle_u | 2 \rangle_l  \; .
\label{state.4.inside}
\end{eqnarray}
%
Obviously the bosonic enhancement leads to the generation of the desired state of the
form~(\ref{best.state.inside}) in 75\% of all cases~\cite{Ou99} whereas the unwanted
contribution $| 2 \rangle_u | 2 \rangle_l$ only occurs in a quarter of all
cases~$(\varepsilon\equiv1)$. Although only an approximation to our goal, this state is readily
available and the desired fourfold reduction of the de Broglie wavelength can be detected using
current technology~\cite{Ou99,ghzexperiment}.
%
%

In order to determine the detectors' response we will now assume that the photons following
channel $u$ are delayed by a tunable phase $\phi$ and are subsequently mixed with the
$l$-channel to form the detector modes $A^\dagger$ and $B^\dagger$:
%
\begin{eqnarray}
u^\dagger = e^{i \phi} \, ({A^\dagger + i \, B^\dagger})/{\sqrt 2}, \quad l^\dagger =
({A^\dagger - i \, B^\dagger})/{\sqrt 2} \, ,
 \label{mode.transformation.2}
\end{eqnarray}
%
see Fig.~\ref{figure1}. For the case of a single photon entering the interferometer through mode
$a^\dagger$ we thus receive the final state
%
\begin{eqnarray}
| \psi \rangle_{out} = \frac{(i+e^{i \phi})| 1 \rangle_A | 0 \rangle_B + (1+i \, e^{i \phi}) | 0
\rangle_A | 1 \rangle_B }{2}
\label{state.out}
\end{eqnarray}
%
and the well known classical photo-detector response probabilities
%
\begin{eqnarray}
P_{A}(\phi) & = & \langle A^\dagger  A  \rangle = \frac{1}{2} \; (1+ \sin \phi)  \\
\mbox{  and  } \quad P_{B}(\phi) & = & \langle B^\dagger B  \rangle = \frac{1}{2} \; (1- \sin
\phi) \; . \label{signal.1}
\end{eqnarray}
%
For the two-photon state~(\ref{state.2.inside}) the corresponding expressions reflect the
halving of the de Broglie wavelength
%
\begin{eqnarray}
P_{AA}(\phi)  =
P_{BB}(\phi) & = &  \frac{1}{4} \; (1 + \cos \, 2\phi)  \\
\mbox{  and  } \quad P_{AB}(\phi) & = &  \frac{1}{2} \; (1 - \cos \, 2\phi) \; .
\label{signal.2}
\end{eqnarray}
%
Here $P_{AA}$, $P_{BB}$ and $P_{AB}\equiv P_{BA}$ stand for the probabilities to detect two
photons in the channels and with the multiplicity indicated by the subscripts. Since it is
difficult to detect single photons and discriminate one from two photons arriving at the same
time, the experimentally most convenient signal is $P_{AB}$.

The four-photon state~(\ref{state.4.inside}) shows the expected reduction to a quarter of the de
Broglie wavelength, namely
%
\begin{eqnarray}
P_{AAAA}(\phi) = P_{BBBB}(\phi) &=&  \frac{ 9 + 12 \cos 2 \phi +
3 \cos 4 \phi }{64} \; , \label{full.signal1.4} \\
P_{AAAB}(\phi) = P_{ABBB}(\phi)& = &  \frac{3 - 3 \, \cos 4 \phi }{16} \; , \label{full.signal2.4}\\
\label{full.signal.4}
\mbox{  and  } \quad P_{AABB}(\phi) & = &  \frac{11 - 12 \cos 2 \phi  + 9
\cos 4 \phi }{32}\; .
%
%
\label{signal.4}
\end{eqnarray}
%
Surprisingly, despite the imperfect form ($\varepsilon = 1$) of the four-photon
state~(\ref{state.4.inside}), $P_{AAAB}$ and $P_{ABBB}$ show a pure fourfold reduction of the
observed de Broglie wavelength with perfect visibility. Here $P$ stands for the four-photon
coincidence probabilities with the channels and their respective detection multiplicity
indicated by the subscripts.

Since it is presently difficult to detect with single photon resolution and discriminate one
from two or more photons arriving at the same time a special detector setup might have to be
used. One can employ multi-port detectors as they are described in~\cite{photonchopping}, a
four-port in channel $A$ suffices to see the signals $P_{AAAA}$ and $P_{AAAB}$ since it can
split up four photons to follow four different channels, for sketches of possible realizations
of four-port detectors see e.g.~\cite{Ou99}.
%
\begin{figure}[htb]
%
\epsfverbosetrue \epsfxsize=3.0in \epsfysize=1.6in
%
\epsffile[0 50 596 715]{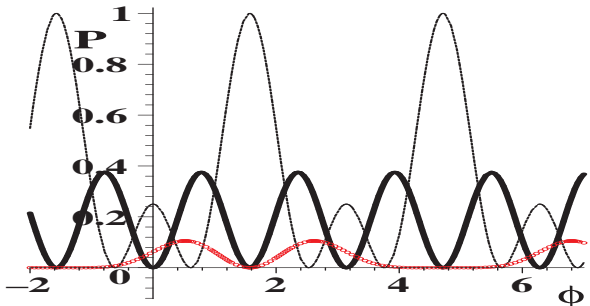}

\caption{\label{figure2} The four-photon coincidence signals as a function of phase
delay~$\phi$: $P_{AAAB}$ (fat line) and $P_{AABB}$ (thin line) in comparison with the classical
signal $P_A P_A P_A P_B$ from \protect{ eq.~(\ref{signal.1})} (dotted red line).}
\end{figure}
%
For photon numbers $2n$ higher than four no pure $2n$-fold wavelength reduction like in
$P_{AAAB}$ is attainable using our scheme; also the problems with the generation and coincidence
detection of more than four photons will make the experiment much harder to perform, we
therefore will not write down the corresponding photo-detection probabilities for 6, 8 or more
photons.
%
%

{\em Speculation:} if one could, however, synthesize an input state of the form
%
\begin{eqnarray}
| \psi \rangle_{in} = \frac{(a^\dagger-b^\dagger)^{2n}+(a^\dagger+b^\dagger)^{2n}}{
\sqrt{2^{2n+1} \cdot 2n!}}  | 0 \rangle \; , \label{best.state.outside}
\end{eqnarray}
%
where $a$ and $b$ stand for any two modes, say polarization, this state would, using the beam
splitter operation~(\ref{mode.transformation}), transform into a "Schr\"odinger-kitten
state"~(\ref{best.state.inside})
%
\begin{eqnarray}
| \psi \rangle_{inside} = \frac{|2n\rangle_u |0\rangle_l + |0\rangle_u |2n \rangle_l}{ \sqrt{2}}
\label{kitten.inside}
\end{eqnarray}
%
and would therefore also yield detector signals with perfect $2n$-fold wavelength reduction for
any photon number $2n$. However, to my knowledge it is not known how to generate such a state
with available technology. Yet, this observation stresses once more the alternatives to
employing non-linear beam splitters~\cite{Jacobson} in order to channel the particles through
the interferometer.
%
%

To conclude, interference patterns of unbound particles with halved de Broglie wavelengths have
been seen in parametric down-conversion. Here a new scheme, using current technology, is
proposed which allows to see further de Broglie wavelength reductions for such unbound particles
below the wavelength of the generating source.

At the level of four photons generated in parametric down-conversion a fourfold reduction for
the interference signal with perfect visibility is achievable.

\acknowledgments I wish to thank Janne Ruostekoski for lively discussions.


\begin{thebibliography}{99}
\bibitem{quartering} Interference patterns with halved de Broglie wavelengths have been seen in parametric
down-conversion, typically these are referred to as interference with the "pump-beam's
wavelength"~\cite{Horne89,RarityTapster,Herzog94}.
\bibitem{lossless} We neglect the effects of detector losses since they are easy to
include~\cite{photonchopping} and do not change our results appreciably but primarily affect the
scheme's success probability.
\end{thebibliography}
\end{document}